\documentclass[runningheads]{llncs}
\usepackage[T1]{fontenc}
\usepackage{graphicx}
\usepackage{amsmath}
\usepackage{amssymb}
\usepackage{subcaption}
\usepackage{xcolor}

\begin{document}
\title{Physics-informed Neural Networks Based on the Fokker-Planck Model for Rarefied Gas Flows}
\titlerunning{PINNs Based on the FP Model for Rarefied Gas Flows}
%
\author{Joonbeom Kim\inst{1}\orcidID{0009-0001-3154-0152} \and
Eunji Jun\inst{1}\orcidID{0000-0001-7403-6784}}
\authorrunning{J. Kim and E. Jun}
%
\institute{
Department of Aerospace Engineering, Korea Advanced Institute of Science and Technology, 34141 Daejeon, Republic of Korea\\
\email{eunji.jun@kaist.ac.kr}}
\maketitle              
%

\begin{abstract}
Rarefied nonequilibrium flows are described by the Boltzmann equation, which can be solved using stochastic particle or deterministic methods. With advances in computing, data-assisted physics-informed neural networks have been widely used for flow prediction by combining governing equations with measured or computed data. Data-free PINNs do not require such training data and have been widely studied for continuum flows, but their application to rarefied flows remains limited. This study develops a data-free Fokker-Planck physics-informed neural network (FP-PINN) to describe rarefied nonequilibrium flows. The physics loss is constructed from the ellipsoidal statistical Fokker-Planck (ES-FP) equation using a velocity distribution function (VDF) closure and moment projection. The FP-PINN is evaluated for a steady one-dimensional Mach 10 normal shock. The FP-PINN results agree with the numerical ES-FP results, with errors below $1\%$. Compared with the particle ES-FP results, the FP-PINN predicts a sharper shock structure with errors up to $17\%$. This difference is attributed to the VDF closure and moment projection.

\end{abstract}

\section{Introduction}


As the gas departs from local thermodynamic equilibrium, the Navier-Stokes equations based on the continuum assumption no longer hold \cite{Bird94_DSMC-book}. The Boltzmann equation instead provides a kinetic description of such flows. The two main classes of numerical methods for solving the Boltzmann equation are stochastic particle methods and deterministic methods. The direct simulation Monte Carlo (DSMC) method is a widely used stochastic particle method that tracks simulation particles through molecular transport and collision steps \cite{Bird94_DSMC-book}. The DSMC method accurately describes nonequilibrium flows but suffers from statistical noise and high computational cost in near-continuum regimes. The discrete velocity method (DVM) is a widely used deterministic method that solves the Boltzmann equation on discrete grids in physical and velocity spaces \cite{Broadwell64_DVM}. The DVM avoids statistical noise but requires large phase-space grids. For both classes of numerical methods, the computational cost of treating collisions can be reduced by using simplified kinetic models such as Fokker-Planck (FP) and Bhatnagar-Gross-Krook (BGK) models \cite{Mathiaud16_ESFP,Kim24_poly,Kim25_USPFPM,Jun18_FPDSMC,Kim25_monamixFP,Kim25_dimixFP,BhatnagarGrossKrook54_BGK,Kim26_UBGK}.


With advances in computing, neural networks have been increasingly used in fluid dynamics. Data-driven neural networks predict flow fields based on experimental or numerical training data. However, their accuracy depends strongly on the amount and distribution of the training data. Raissi et al. introduced physics-informed neural networks (PINNs) \cite{Raissi19_PINNs}, which incorporate residuals of the governing equations into the loss function to provide physical information during training. Data-assisted PINNs combine these residuals with measured or computed data to reconstruct flow properties and estimate unknown physical parameters. However, data-assisted PINNs still require measured or computed data, which can be costly to obtain.


Alongside data-assisted PINNs, data-free PINNs have been studied for flow prediction without measured or computed training data. Their loss functions are constructed from residuals associated with the governing equations, boundary conditions, and initial conditions. Data-free PINNs have been widely studied for continuum flows governed by the Euler and Navier-Stokes equations, including shocks. In comparison, relatively few studies have considered rarefied gas flows. Previous studies focused mainly on low-speed or near-equilibrium rarefied gas flows governed by the BGK equation \cite{Zhang23_PINNDVM}. More recently, Oh et al. used separable PINNs to solve the BGK equation for transient smooth and Riemann problems, but their validation focused on density, velocity, and temperature \cite{Oh25_SPINNs}. Since stress and heat flux characterize nonequilibrium momentum and energy transport in rarefied gas flows, data-free PINNs should accurately predict these quantities to describe rarefied nonequilibrium flows.


This paper develops a data-free Fokker-Planck physics-informed neural network (FP-PINN) to describe rarefied nonequilibrium flows by predicting density, bulk velocity, temperature, stress, and heat flux. The physics loss is constructed from the ellipsoidal statistical Fokker-Planck (ES-FP) equation using a velocity distribution function (VDF) closure and moment projection to avoid velocity-space discretization. The remainder of the paper is organized as follows. Section \ref{sec:FP} describes the ES-FP model, VDF closure, and moment projection. Section \ref{sec:FP-PINN} presents the FP-PINN architecture, loss functions, and training settings. Section \ref{sec:results} presents results for a steady one-dimensional Mach 10 normal shock. Section \ref{sec:conclusions} concludes the paper.

\section{\label{sec:FP}Fokker-Planck model}


The ES-FP model describes the evolution of the VDF $f(\boldsymbol{x},\boldsymbol{c},t)$ through drift and diffusion in velocity space, where $\boldsymbol{x}$, $\boldsymbol{c}$, and $t$ denote position, molecular velocity, and time, respectively \cite{Mathiaud16_ESFP}. To express the ES-FP equation in nondimensional form, the reference state is specified by the temperature $T_{\mathrm{ref}}$ and number density $n_{\mathrm{ref}}$. The reference mass density $\rho_{\mathrm{ref}}$, length $L_{\mathrm{ref}}$, velocity $V_{\mathrm{ref}}$, and time $t_{\mathrm{ref}}$ are calculated as
\begin{equation}
\begin{aligned}
    \rho_{\mathrm{ref}} = mn_{\mathrm{ref}},                             
    \quad
    L_{\mathrm{ref}} = \lambda_{\mathrm{ref}} = \frac{1}{\sqrt{2}\pi d^2n_{\mathrm{ref}}},                      
    \quad
    V_{\mathrm{ref}}    = \sqrt{RT_{\mathrm{ref}}},                         
    \quad                 
    t_{\mathrm{ref}}    = \frac{L_{\mathrm{ref}}}{V_{\mathrm{ref}}},
\end{aligned}
    \label{eq:reference-scales}
\end{equation}
where $m$ is the molecular mass, $\lambda_{\mathrm{ref}}$ is the reference mean free path, $d$ is the molecular diameter, $R=k_B/m$ is the specific gas constant, and $k_B$ is the Boltzmann constant. Using these reference scales, the dimensional variables are nondimensionalized as
\begin{equation}
\begin{aligned}
    \hat{x}_i &= \frac{x_i}{L_{\mathrm{ref}}}, &
    \hat{t} &= \frac{t}{t_{\mathrm{ref}}}, &
    \hat{c}_i &= \frac{c_i}{V_{\mathrm{ref}}}, &
    \hat{\rho} &= \frac{\rho}{\rho_{\mathrm{ref}}}, &
    \hat{U}_i &= \frac{U_i}{V_{\mathrm{ref}}}, \\
    \hat{T} &= \frac{T}{T_{\mathrm{ref}}}, &
    \hat{p} &= \frac{p}{\rho_{\mathrm{ref}}V_{\mathrm{ref}}^2}, &
    \hat{\sigma}_{ij} &= \frac{\sigma_{ij}}{\rho_{\mathrm{ref}}V_{\mathrm{ref}}^2}, &
    \hat{q}_i &= \frac{q_i}{\rho_{\mathrm{ref}}V_{\mathrm{ref}}^3}, &
    \hat{f} &= \frac{f}{\rho_{\mathrm{ref}}V_{\mathrm{ref}}^{-3}},
\end{aligned}
\label{eq:nondimensional-variables}
\end{equation}
where a hat denotes a nondimensional variable, $U_i$ is the bulk velocity, $p=\rho RT$ is the pressure, $\sigma_{ij}$ is the deviatoric stress tensor, and $q_i$ is the heat-flux vector. The nondimensional ES-FP equation is written in residual form as
\begin{equation}
\begin{aligned}
    \mathcal{R}_{\mathrm{FP}}
    (\hat{\boldsymbol{x}},\hat{\boldsymbol{c}},\hat{t})
    =
    \frac{\partial \hat{f}}{\partial \hat{t}}
    +\hat{c}_i\frac{\partial \hat{f}}{\partial \hat{x}_i}
    -\frac{1}{\hat{\tau}}
    \frac{\partial}{\partial \hat{c}_i}
    \left(\hat{C}_i\hat{f}\right)
    -\frac{1}{\hat{\tau}}
    \frac{\partial^2}{\partial \hat{c}_i\partial \hat{c}_j}
    \left[
        \left(
            \hat{T}\delta_{ij}
            +\frac{\beta\hat{\sigma}_{ij}}{\hat{\rho}}
        \right)\hat{f}
    \right]
    =0,
    \label{eq:esfp}
\end{aligned}
\end{equation}
where $\hat{C}_i=\hat{c}_i-\hat{U}_i$ is the nondimensional thermal velocity, $\delta_{ij}$ is the Kronecker delta, $\hat{\tau}$ is the nondimensional relaxation time, and $\beta$ is the ES-FP parameter. The Einstein summation convention is used throughout this paper. The ES-FP parameter is defined as
\begin{equation}
    \beta
    =
    \max
    \left[
        1-\frac{3}{2\mathrm{Pr}},
        -\frac{\hat{\rho}\hat{T}}
        {\lambda_{\max}-\hat{\rho}\hat{T}}
    \right],
    \label{eq:beta}
\end{equation}
where $\mathrm{Pr}=2/3$ is the Prandtl number for a monatomic gas and $\lambda_{\max}$ is the maximum eigenvalue of the nondimensional pressure tensor $\hat{P}_{ij}=\hat{\rho}\hat{T}\delta_{ij}+\hat{\sigma}_{ij}$. The viscosity $\mu = \mu_{\mathrm{ref}}\hat{T}^{\omega}$ is evaluated using the variable soft sphere (VSS) model \cite{Bird94_DSMC-book}, where $\mu_{\mathrm{ref}}$ is the reference viscosity, $T_{\mathrm{VSS}}$ is the VSS reference temperature, and $\omega$ is the viscosity index. The nondimensional relaxation time is
\begin{equation}
    \hat{\tau}
    =
    \frac{\tau}{t_{\mathrm{ref}}}
    =
    \frac{2(1-\beta)\mu/p}{t_{\mathrm{ref}}}
    =
    \frac{2(1-\beta)\mu_{\mathrm{ref}}}
    {t_{\mathrm{ref}}\rho_{\mathrm{ref}}V_{\mathrm{ref}}^2}
    \frac{\hat{T}^{\omega-1}}{\hat{\rho}}.
    \label{eq:relaxation-time}
\end{equation}


A finite set of macroscopic moments does not uniquely determine the VDF. A VDF closure is therefore required to approximate the VDF using these moments. This study uses the Grad 26-moment distribution as the VDF closure, given by
\begin{equation}
\begin{aligned}
    \hat{f}_{G26}
    =
    \hat{f}_M
    \Bigg[
        &1
        +\frac{\hat{\sigma}_{ij}\hat{C}_i\hat{C}_j}
        {2\hat{\rho}\hat{T}^2}
        +\frac{\hat{q}_i\hat{C}_i}
        {5\hat{\rho}\hat{T}^2}
        \left(
            \frac{\hat{C}^2}{\hat{T}}-5
        \right)
        +\frac{\hat{m}_{ijk}\hat{C}_i\hat{C}_j\hat{C}_k}
        {6\hat{\rho}\hat{T}^3}
        \\
        &+\frac{\hat{R}_{ij}\hat{C}_i\hat{C}_j}
        {28\hat{\rho}\hat{T}^3}
        \left(
            \frac{\hat{C}^2}{\hat{T}}-7
        \right)
        +\frac{\hat{\Delta}}
        {120\hat{\rho}\hat{T}^2}
        \left(
            \frac{\hat{C}^4}{\hat{T}^2}
            -10\frac{\hat{C}^2}{\hat{T}}
            +15
        \right)
    \Bigg]
    ,
    \label{eq:g26}
\end{aligned}
\end{equation}
where $\hat{f}_M=
    \hat{\rho} (2\pi\hat{T})^{-3/2} \exp( -\hat{C}^2/{2\hat{T}} )$ is the Maxwell distribution.
The higher-order moments in Eq.~\eqref{eq:g26} are $m_{ijk}$, $R_{ij}$, and $\Delta$, with $\hat{m}_{ijk}=m_{ijk}/(\rho_{\mathrm{ref}}V_{\mathrm{ref}}^3)$, $\hat{R}_{ij}=R_{ij}/(\rho_{\mathrm{ref}}V_{\mathrm{ref}}^4)$, and $\hat{\Delta}=\Delta/(\rho_{\mathrm{ref}}V_{\mathrm{ref}}^4)$. These moments are determined from $(\rho,U_i,T,\sigma_{ij},q_i)$ and their spatial derivatives using the R13 constitutive relations~\cite{Torrilhon04_R13}. For a steady one-dimensional normal shock, the required higher-order moments are
\begin{equation}
\begin{aligned}
    \hat{m}_{xxx}
    =
    -\frac{6}{5}
    \frac{\hat{T}^{\omega-1}}{\hat{\rho}}
    \Bigg[
        &\hat{T}
        \frac{\partial\hat{\sigma}_{xx}}{\partial\hat{x}}
        -\frac{\hat{T}\hat{\sigma}_{xx}}{\hat{\rho}}
        \frac{\partial\hat{\rho}}{\partial\hat{x}}
        +\frac{8\hat{q}_x}{15}
        \frac{\partial\hat{U}_x}{\partial\hat{x}}
        -\frac{\hat{\sigma}_{xx}}{\hat{\rho}}
        \frac{\partial\hat{\sigma}_{xx}}{\partial\hat{x}}
    \Bigg]
    ,
    \label{eq:r13-mxxx}
\end{aligned}
\end{equation}
\begin{equation}
\begin{aligned}
    \hat{R}_{xx}
    =
    -\frac{24}{5}
    \frac{\hat{T}^{\omega-1}}{\hat{\rho}}
    \Bigg[
        \Biggl(
            \frac{2\hat{T}}{3}
            &-\frac{5\hat{\sigma}_{xx}}{6\hat{\rho}}
        \Biggr)
        \frac{\partial\hat{q}_x}{\partial\hat{x}}
        -\frac{2\hat{q}_x}{3\hat{\rho}}
        \frac{\partial\hat{\sigma}_{xx}}{\partial\hat{x}}
        +\frac{2\hat{q}_x}{3}
        \frac{\partial\hat{T}}{\partial\hat{x}}
        \\
        &-\frac{2\hat{T}\hat{q}_x}{3\hat{\rho}}
        \frac{\partial\hat{\rho}}{\partial\hat{x}}
        +\left(
            \frac{10\hat{T}\hat{\sigma}_{xx}}{21}
            -\frac{5\hat{\sigma}_{xx}^2}{6\hat{\rho}}
        \right)
        \frac{\partial\hat{U}_x}{\partial\hat{x}}
    \Bigg]
    ,
    \label{eq:r13-Rxx}
\end{aligned}
\end{equation}
\begin{equation}
\begin{aligned}
    \hat{\Delta}
    =
    -12
    \frac{\hat{T}^{\omega-1}}{\hat{\rho}}
    \Bigg[
        &\hat{T}
        \frac{\partial\hat{q}_x}{\partial\hat{x}}
        +\frac{5\hat{q}_x}{2}
        \frac{\partial\hat{T}}{\partial\hat{x}}
        -\frac{\hat{T}\hat{q}_x}{\hat{\rho}}
        \frac{\partial\hat{\rho}}{\partial\hat{x}}
        -\frac{\hat{q}_x}{\hat{\rho}}
        \frac{\partial\hat{\sigma}_{xx}}{\partial\hat{x}}
        +\hat{T}\hat{\sigma}_{xx}
        \frac{\partial\hat{U}_x}{\partial\hat{x}}
    \Bigg]
    .
    \label{eq:r13-Delta}
\end{aligned}
\end{equation}


The ES-FP equation still depends on molecular velocity after the VDF closure is applied. To eliminate this velocity-space dependence, the ES-FP residual $\mathcal{R}_{\mathrm{FP}}$ in Eq.~\eqref{eq:esfp} is projected onto the velocity-moment basis functions $\psi_k$ as
\begin{equation}
\begin{aligned}
    \mathcal{R}_{\psi_k}
    =
    \int_{\mathbb{R}^3}
    \psi_k\mathcal{R}_{\mathrm{FP}}
    \,d\hat{\boldsymbol{c}}
    =0,
    \quad
    \psi_k
    \in
    \left\{
        1,
        \hat{C}_i,
        \hat{C}^2-3\hat{T},
        \hat{C}_{\langle i}\hat{C}_{j\rangle},
        (\hat{C}^2-5\hat{T})\hat{C}_i
    \right\}
    ,
    \label{eq:moment-projection}
\end{aligned}
\end{equation}
where the angular brackets denote the symmetric trace-free part. For a steady one-dimensional normal shock, the projected equations reduce to
\begin{equation}
    \mathcal{R}_{\rho}
    =\hat{U}_x\frac{\partial\hat{\rho}}{\partial\hat{x}}
    +\hat{\rho}\frac{\partial\hat{U}_x}{\partial\hat{x}}
    =0,
    \label{eq:projected-mass}
\end{equation}
\begin{equation}
    \mathcal{R}_{U_x}
    =\hat{T}\frac{\partial\hat{\rho}}{\partial\hat{x}}
    +\hat{\rho}\frac{\partial\hat{T}}{\partial\hat{x}}
    +\hat{\rho}\hat{U}_x
    \frac{\partial\hat{U}_x}{\partial\hat{x}}
    +\frac{\partial\hat{\sigma}_{xx}}{\partial\hat{x}}
    =0,
    \label{eq:projected-momentum}
\end{equation}
\begin{equation}
\begin{aligned}
    \mathcal{R}_T
    =&2\hat{\rho}\hat{T}
    \frac{\partial\hat{U}_x}{\partial\hat{x}}
    +3\hat{\rho}\hat{U}_x
    \frac{\partial\hat{T}}{\partial\hat{x}}
    +2\frac{\partial\hat{q}_x}{\partial\hat{x}}
    +2\hat{\sigma}_{xx}
    \frac{\partial\hat{U}_x}{\partial\hat{x}}
    =0,
    \label{eq:projected-energy}
\end{aligned}
\end{equation}
\begin{equation}
\begin{aligned}
    \mathcal{R}_{\sigma_{xx}}
    =\frac{4}{3}\hat{\rho}\hat{T}
    \frac{\partial\hat{U}_x}{\partial\hat{x}}
    +\frac{2(1-\beta)}{\hat{\tau}}\hat{\sigma}_{xx}
    &+\frac{\partial\hat{m}_{xxx}}{\partial\hat{x}}
    +\frac{8}{15}\frac{\partial\hat{q}_x}{\partial\hat{x}}
    \\
    &+\frac{7}{3}\hat{\sigma}_{xx}
    \frac{\partial\hat{U}_x}{\partial\hat{x}}
    +\hat{U}_x
    \frac{\partial\hat{\sigma}_{xx}}{\partial\hat{x}}
    =0,
    \label{eq:projected-stress}
\end{aligned}
\end{equation}
\begin{equation}
\begin{aligned}
    \mathcal{R}_{q_x}
    =\frac{1}{3}\frac{\partial\hat{\Delta}}{\partial\hat{x}}
    &+\frac{\partial\hat{R}_{xx}}{\partial\hat{x}}
    +5\hat{\rho}\hat{T}
    \frac{\partial\hat{T}}{\partial\hat{x}}
    +2\hat{T}
    \frac{\partial\hat{\sigma}_{xx}}{\partial\hat{x}}
    +7\hat{\sigma}_{xx}
    \frac{\partial\hat{T}}{\partial\hat{x}}
    +\frac{6}{\hat{\tau}}\hat{q}_x
    \\
    &+2\hat{m}_{xxx}
    \frac{\partial\hat{U}_x}{\partial\hat{x}}
    +\frac{32}{5}\hat{q}_x
    \frac{\partial\hat{U}_x}{\partial\hat{x}}
    +2\hat{U}_x
    \frac{\partial\hat{q}_x}{\partial\hat{x}}
    +2\hat{\sigma}_{xx}\hat{U}_x
    \frac{\partial\hat{U}_x}{\partial\hat{x}}
    =0.
    \label{eq:projected-heat}
\end{aligned}
\end{equation}

\section{\label{sec:FP-PINN}Fokker-Planck Physics-informed Neural Network}


The FP-PINN consists of a neural network and loss functions constructed from the projected ES-FP equations and prescribed constraints. For a steady one-dimensional normal shock, the network maps the spatial coordinate $\hat{x}$ to the moments $(\hat{\rho},\hat{U}_x,\hat{T},\hat{\sigma}_{xx},\hat{q}_x)$.
The FP-PINN uses a fully connected neural network with four hidden layers and 40 nodes in each layer. The hyperbolic tangent activation function is applied after each hidden layer. The nondimensional coordinate $\hat{x}\in[0,\hat{L}]$ is mapped to the network input $\xi=2\hat{x}/\hat{L}-1\in[-1,1]$, where $L$ is the domain length and $\hat{L}=L/L_{\mathrm{ref}}$. The network outputs $(z_{\rho},\hat{U}_x,z_T,\hat{\sigma}_{xx},\hat{q}_x)$. The softplus function, $\operatorname{softplus}(z)=\ln[1+\exp(z)]$, is applied as $\hat{\rho}=\operatorname{softplus}(z_{\rho})$ and $
  \hat{T}=\operatorname{softplus}(z_T)$ to enforce $\hat{\rho}>0$ and $\hat{T}>0$.


The FP-PINN is trained by minimizing a total loss $L_{\mathrm{total}} = L_{\mathrm{phys}} +L_{\mathrm{BC}}+L_{\mathrm{center}}$, where $L_{\mathrm{phys}}$, $L_{\mathrm{BC}}$, and $L_{\mathrm{center}}$ denote the physics, boundary, and center losses, respectively. The physics loss is based on the five projected residuals defined in Eqs.~\eqref{eq:projected-mass}-\eqref{eq:projected-heat}. To limit loss reduction caused by decreases in the predicted density or temperature, each residual is divided by its local scale as
\begin{equation}
\begin{aligned}
    \widetilde{\mathcal{R}}_{\rho}
    =
    \frac{\mathcal{R}_{\rho}}{\hat{\rho}},
    \quad
    \widetilde{\mathcal{R}}_{U_x}
    =
    \frac{\mathcal{R}_{U_x}}{\hat{\rho}\hat{T}},
    \quad
    \widetilde{\mathcal{R}}_{T}
    =
    \frac{\mathcal{R}_{T}}{\hat{\rho}\hat{T}},
    \quad
    \widetilde{\mathcal{R}}_{\sigma_{xx}}
    =
    \frac{\mathcal{R}_{\sigma_{xx}}}{\hat{\rho}\hat{T}},
    \quad
    \widetilde{\mathcal{R}}_{q_x}
    =
    \frac{\mathcal{R}_{q_x}}{\hat{\rho}\hat{T}^2}
    .
\end{aligned}
    \label{eq:scaled-residuals}
\end{equation}
The physics loss is then constructed as $L_{\mathrm{phys}}=\langle\sum_{j\in\{\rho,U_x,T,\sigma_{xx},q_x\}}\widetilde{\mathcal{R}}_j^2\rangle_{\Omega_{\mathrm{phys}}}$, where $\Omega_{\mathrm{phys}}$ denotes the set of physics collocation points and $\langle\cdot\rangle_{\Omega_{\mathrm{phys}}}$ denotes the mean over $\Omega_{\mathrm{phys}}$. With $\mathcal{M}=\{\hat{\rho},\hat{U}_x,\hat{T},\hat{\sigma}_{xx},\hat{q}_x\}$, the boundary loss is $L_{\mathrm{BC}}=\sum_{\mathrm{bc}\in\{\mathrm{up},\mathrm{down}\}}\langle\sum_{\phi\in\mathcal{M}} (\phi-\phi_{\mathrm{bc}})^2\rangle_{\Omega_{\mathrm{bc}}}$, where $\Omega_{\mathrm{bc}}$ denotes the collocation points in boundary region $\mathrm{bc}$ and $\phi_{\mathrm{bc}}$ denotes the prescribed value of $\phi$ in that region. The projected equations do not fix the shock position. The center loss is therefore defined as $L_{\mathrm{center}}=[\hat{U}_{x,\mathrm{center}}-(\hat{U}_{x,\mathrm{up}}+\hat{U}_{x,\mathrm{down}})/2]^2$, where $\hat{U}_{x,\mathrm{center}}$ is the velocity at the domain center and $\hat{U}_{x,\mathrm{up}}$ and $\hat{U}_{x,\mathrm{down}}$ are the prescribed upstream and downstream velocities, respectively.


\begin{figure}[!b]
    \centering
    \includegraphics[width=0.9\linewidth]{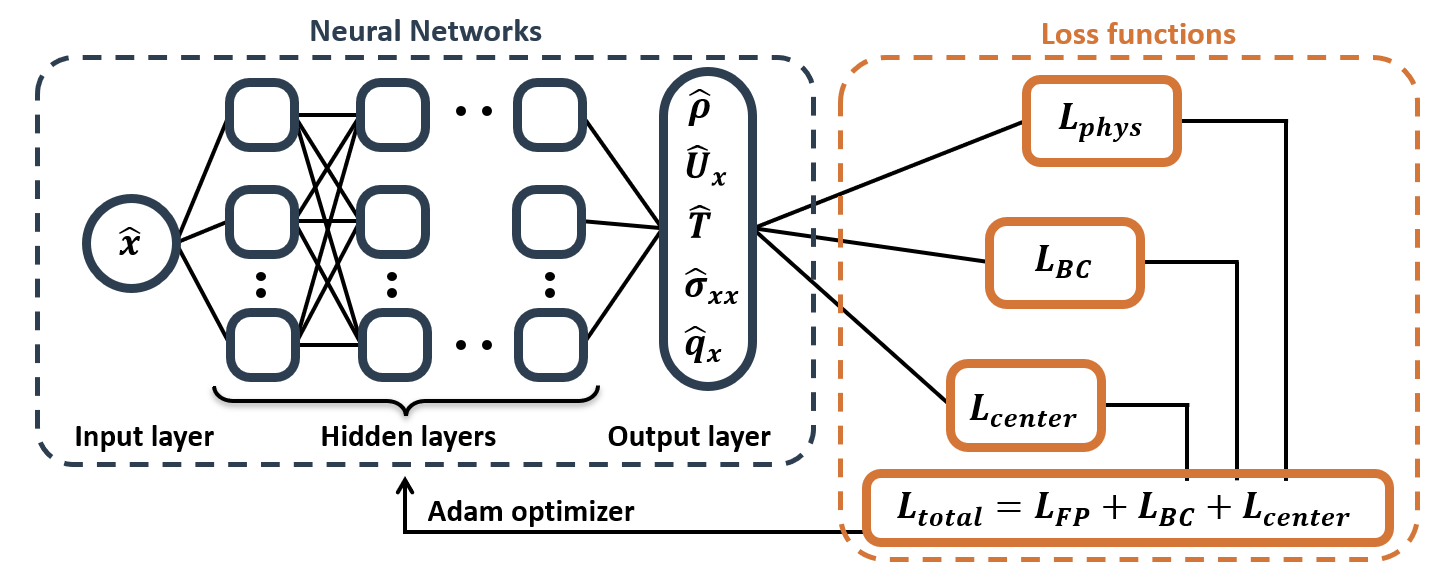}
    \caption{Overall structure of the FP-PINN.}
    \label{fig:FP-PINN}
\end{figure}

The network weights are initialized using the Xavier uniform method \cite{GlorotBengio10_Xavier}. The hidden-layer biases are initialized to zero, and the output-layer biases are set to represent the upstream equilibrium state. The network is trained using the Adam optimizer with a fixed learning rate of $10^{-3}$ \cite{KingmaBa15_Adam}. The FP-PINN is implemented in PyTorch for ease of development and debugging \cite{PaszkeEtAl19_PyTorch}. The overall structure of the FP-PINN is summarized in Fig.~\ref{fig:FP-PINN}.

\section{\label{sec:results}Results}

\begin{figure}[!b]
	\centering
	\begin{subfigure}{0.32\textwidth}
		\centering
		\includegraphics[width=1.\linewidth]{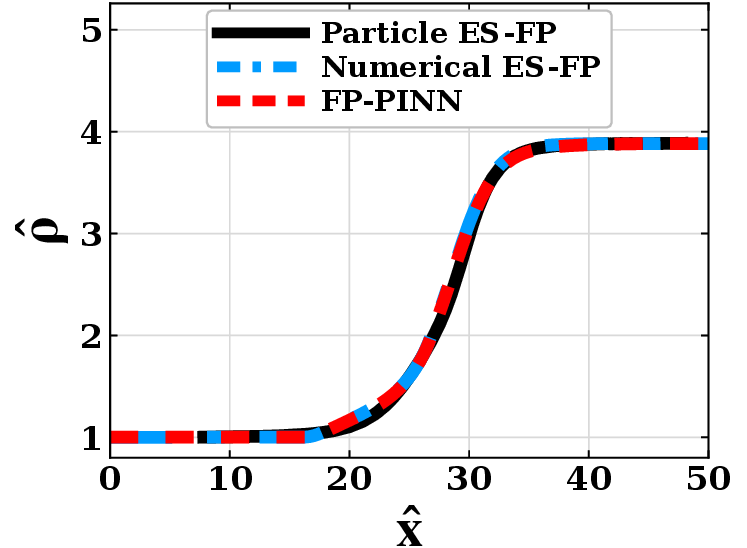}
            \caption{\label{subfig:density}Density.}
	\end{subfigure}
	\begin{subfigure}{0.32\textwidth}
		\centering
		\includegraphics[width=1.\linewidth]{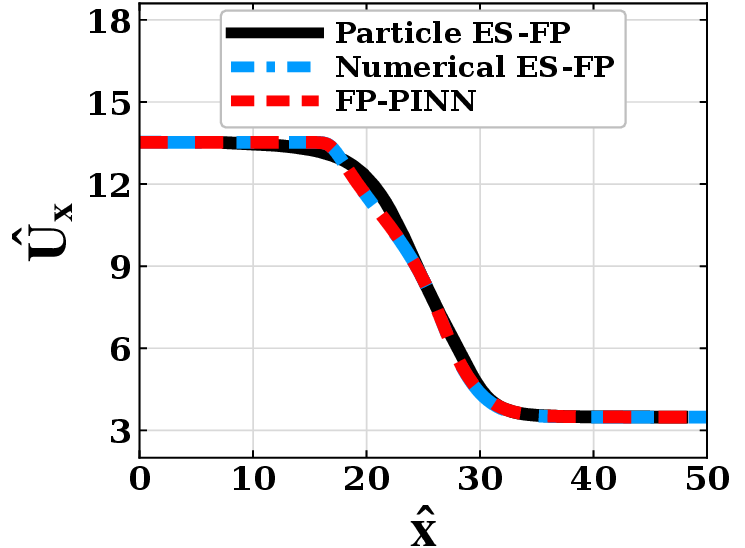}
            \caption{\label{subfig:velocity}Bulk velocity.}
	\end{subfigure}
	\begin{subfigure}{0.32\textwidth}
		\centering
		\includegraphics[width=1.\linewidth]{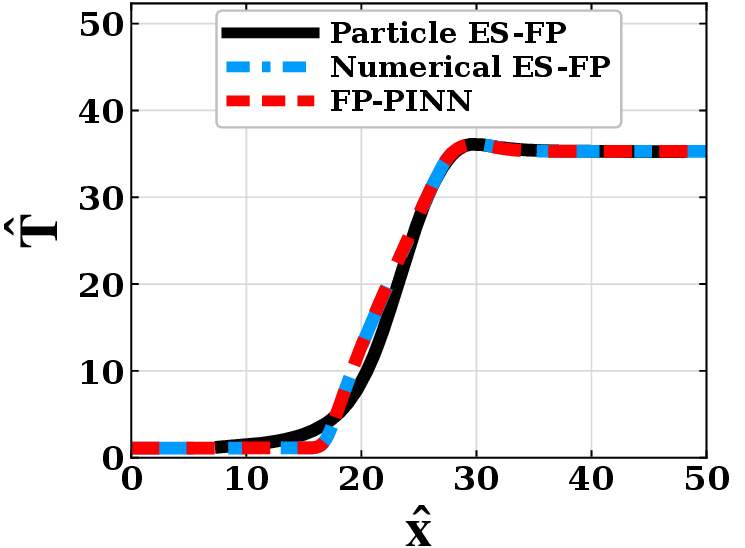}
            \caption{\label{subfig:temperature}Temperature.}
	\end{subfigure}
	\centering
	\begin{subfigure}{0.32\textwidth}
		\centering
		\includegraphics[width=1.\linewidth]{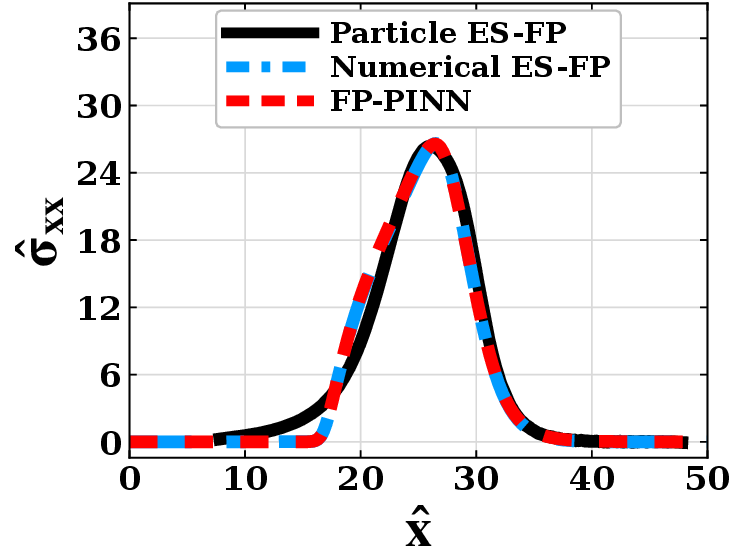}
            \caption{\label{subfig:stress}Normal stress.}
	\end{subfigure}
	\begin{subfigure}{0.32\textwidth}
		\centering
		\includegraphics[width=1.\linewidth]{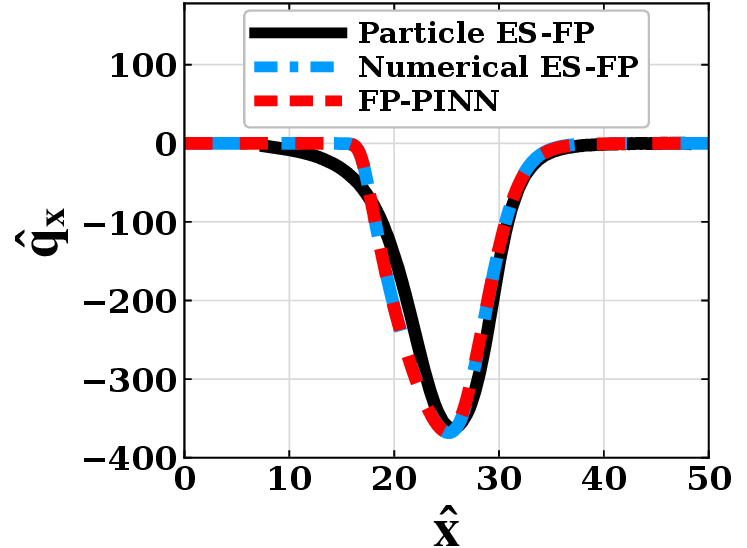}
            \caption{\label{subfig:heatflux}Heat flux.}
	\end{subfigure}
    \caption{\label{fig:shock}Moment profiles for the Mach 10 normal shock.}
\end{figure}

\begin{table}[!b]
\caption{\label{tab:L2norm}Normalized L2 errors of FP-PINN results with respect to the particle ES-FP and numerical ES-FP results.}
    \begin{tabular}{|l|ccccc|}
        \hline
            \rule{0pt}{2.7ex}~Reference &
            ~Density &
            ~Bulk velocity &
            ~Temperature &
            ~Normal stress &
            ~Heat flux ~ \\[0.5ex]
        \hline 
            \rule{0pt}{2.8ex}~Particle ES-FP~  & 1.51\% & 2.29\% & 5.14\% & 13.20\% & 17.00\% \\[0.6ex]
            ~Numerical ES-FP~ & 0.54\% & 0.14\% & 0.22\% & \phantom{1}0.97\% & \phantom{1}0.65\% \\[0.5ex]
        \hline
    \end{tabular}
\end{table}


A steady one-dimensional Mach 10 normal shock is considered to assess the FP-PINN. Argon is used as the working gas with a molecular mass of $m=6.63\times10^{-26}\,\mathrm{kg}$. The VSS reference diameter, reference temperature, viscosity index, and scattering parameter for argon are $d_{\mathrm{VSS}}=4.11\times10^{-10}\,\mathrm{m}$, $T_{\mathrm{VSS}}=273.15\,\mathrm{K}$, $\omega=0.81$, and $\alpha=1.4$, respectively. The upstream number density and temperature are $n_{\mathrm{up}}=1.6095\times10^{21}\,\mathrm{m}^{-3}$ and $T_{\mathrm{up}}=300\,\mathrm{K}$, respectively \cite{Fei20_comparison}. Using $n_{\mathrm{ref}}=n_{\mathrm{up}}$ and $T_{\mathrm{ref}}=T_{\mathrm{VSS}}$, the nondimensional upstream state is $(\hat{\rho}_{\mathrm{up}},\allowbreak \hat{U}_{x,\mathrm{up}},\allowbreak \hat{T}_{\mathrm{up}},\allowbreak \hat{\sigma}_{xx,\mathrm{up}},\allowbreak \hat{q}_{x,\mathrm{up}})=\allowbreak (1,\allowbreak 13.5296,\allowbreak 1.09830,\allowbreak 0,\allowbreak 0)$. The nondimensional downstream state obtained from the Rankine-Hugoniot relations is $(\hat{\rho}_{\mathrm{down}},\allowbreak \hat{U}_{x,\mathrm{down}},\allowbreak \hat{T}_{\mathrm{down}},\allowbreak \hat{\sigma}_{xx,\mathrm{down}},\allowbreak \hat{q}_{x,\mathrm{down}})=\allowbreak (3.88350,\allowbreak 3.48387,\allowbreak 35.2808,\allowbreak 0,\allowbreak 0)$. The computational domain spans $\hat{x}\in[0,50]$. The physics loss uses 1000 uniformly spaced collocation points over $[0,50]$. The boundary loss uses 10 uniformly spaced collocation points in each boundary region, with $\hat{x}\in[0,5]$ for the upstream region and $\hat{x}\in[45,50]$ for the downstream region. The center loss is evaluated at $\hat{x}=25$. The FP-PINN results are compared with the particle ES-FP results and the numerical ES-FP results. The particle ES-FP results are obtained using 1000 cells, approximately 10000 simulation particles per upstream cell, and a time step of $10^{-9}\,\mathrm{s}$. The numerical ES-FP results are obtained by solving the projected ES-FP equations in Eqs.~\eqref{eq:projected-mass}-\eqref{eq:projected-heat} using the Newton method with 1200 interior points and two prescribed boundary points \cite{Torrilhon04_R13}.


Figures~\ref{subfig:density} to~\ref{subfig:heatflux} show density, bulk velocity, temperature, normal stress, and heat flux profiles for the Mach 10 normal shock, respectively. The black solid, blue dashed, and red dashed lines represent the particle ES-FP, numerical ES-FP, and FP-PINN results, respectively. Across the shock layer, density and temperature increase, while bulk velocity decreases. The normal stress and heat flux attain their peak magnitudes within the shock layer. The differences between the FP-PINN results and the particle and numerical ES-FP results are quantified using the relative L2 error. For each moment $\phi$, the relative $L_2$ error is defined as $E_{L_2}(\phi_{\mathrm{FP\text{-}PINN}})=\|\phi_{\mathrm{FP\text{-}PINN}}-\phi_{\mathrm{ref}}\|_2/\|\phi_{\mathrm{ref}}\|_2$, where $\phi_{\mathrm{ref}}$ denotes the particle ES-FP or numerical ES-FP result. The calculated errors are summarized in Table~\ref{tab:L2norm}. The FP-PINN results have good agreement with the numerical ES-FP results, with errors below $1\%$. However, the FP-PINN results exhibit a sharper shock structure than the particle ES-FP results, with errors up to $17\%$. These differences are associated with the Grad 26-moment VDF closure and truncation of the projected ES-FP equation at the heat flux moment.

\section{\label{sec:conclusions}Conclusions}


This study developed a data-free FP-PINN to predict moments in rarefied nonequilibrium flows using physics losses derived from the ES-FP equation, a VDF closure, and moment projection. For a steady one-dimensional Mach 10 normal shock, the FP-PINN agreed with the numerical ES-FP results within $1\%$ but predicted a sharper shock than the particle ES-FP results, with errors up to $17\%$. This difference is attributed to the Grad 26-moment VDF closure and projection up to the heat-flux. Future works include additional higher-order moments.

\begin{credits}
\subsubsection{\ackname} This work was supported by the National Research Foundation of Korea(NRF) grant funded by the Korea government(MSIT) (RS-2025-00557012). 

\subsubsection{\discintname}
The authors have no competing interests to declare that are
relevant to the content of this article.
\end{credits}
%
%
%

\begin{thebibliography}{8}


\bibitem{Bird94_DSMC-book}
Bird, G.A.: Molecular Gas Dynamics and the Direct Simulation of Gas Flows. Oxford University Press, Oxford (1994). \doi{10.1093/oso/9780198561958.001.0001}

\bibitem{Broadwell64_DVM}
Broadwell, J.E.: Study of rarefied shear flow by the discrete velocity method. J. Fluid Mech. \textbf{19}(3), 401--414 (1964). \doi{10.1017/S0022112064000817}

\bibitem{Mathiaud16_ESFP}
Mathiaud, J., Mieussens, L.: A Fokker--Planck model of the Boltzmann equation with correct Prandtl number. J. Stat. Phys. \textbf{162}(2), 397--414 (2016). \doi{10.1007/s10955-015-1404-9}

\bibitem{Kim24_poly}
Kim, J., Kim, S., Jun, E.: A spatial interpolation for a stochastic particle Fokker--Planck model using a polynomial reconstruction. Phys. Fluids \textbf{36}(12), 127171 (2024). \doi{10.1063/5.0247494}

\bibitem{Kim25_USPFPM}
Kim, J., Jun, E.: A second-order particle Fokker--Planck--Master method for diatomic gas flows. Phys. Fluids \textbf{37}(11), 117117 (2025). \doi{10.1063/5.0292414}

\bibitem{Jun18_FPDSMC}
Jun, E., Gorji, M.H., Grabe, M., Hannemann, K.: Assessment of the cubic Fokker–Planck–DSMC hybrid method for hypersonic rarefied flows past a cylinder. Comput. Fluids \textbf{168}, 1--13 (2018). \doi{10.1016/j.compfluid.2018.03.059}

\bibitem{Kim25_monamixFP}
Kim, S., Jun, E.: A particle Fokker–Planck method for rarefied gas flows of monatomic mixtures. Phys. Fluids \textbf{37}(1), 016104 (2025). \doi{10.1063/5.0247155}

\bibitem{Kim25_dimixFP}
Kim, S., Jun, E.: A stochastic particle method based on the Fokker–Planck Master equation for rarefied gas flows of diatomic mixtures. Phys. Fluids \textbf{37}(3), 036111 (2025). \doi{10.1063/5.0258301}

\bibitem{BhatnagarGrossKrook54_BGK}
Bhatnagar, P.L., Gross, E.P., Krook, M.: A model for collision processes in gases. I. Small amplitude processes in charged and neutral one-component systems. Phys. Rev. \textbf{94}(3), 511--525 (1954). \doi{10.1103/PhysRev.94.511}

\bibitem{Kim26_UBGK}
Kim, I., Kim, J., Park, W., Jun, E.: A particle multi-relaxation Bhatnagar--Gross--Krook method for rarefied monatomic gas mixtures. Phys. Fluids \textbf{38}(8), 087001 (2026). \doi{10.1063/5.0340698}

\bibitem{Raissi19_PINNs}
Raissi, M., Perdikaris, P., Karniadakis, G.E.: Physics-informed neural networks: A deep learning framework for solving forward and inverse problems involving nonlinear partial differential equations. J. Comput. Phys. \textbf{378}, 686--707 (2019). \doi{10.1016/j.jcp.2018.10.045}

\bibitem{Zhang23_PINNDVM}
Zhang, L., Ma, W., Lou, Q., Zhang, J.: Simulation of rarefied gas flows using physics-informed neural network combined with discrete velocity method. Phys. Fluids \textbf{35}(7), 077124 (2023). \doi{10.1063/5.0156404}

\bibitem{Oh25_SPINNs}
Oh, J., Cho, S.Y., Yun, S.-B., Park, E., Hong, Y.: Separable physics-informed neural networks for solving the BGK model of the Boltzmann equation. SIAM J. Sci. Comput. \textbf{47}(2), C451--C474 (2025). \doi{10.1137/24M1668809}

\bibitem{Fei20_comparison}
Fei, F., Liu, H., Liu, Z., Zhang, J.: A benchmark study of kinetic models for shock waves. AIAA J. \textbf{58}(6), 2596--2608 (2020). \doi{10.2514/1.J059029}

\bibitem{Torrilhon04_R13}
Torrilhon, M., Struchtrup, H.: Regularized 13-moment equations: Shock structure calculations and comparison to Burnett models. J. Fluid Mech. \textbf{513}, 171--198 (2004). \doi{10.1017/S0022112004009917}

\bibitem{GlorotBengio10_Xavier}
Glorot, X., Bengio, Y.: Understanding the difficulty of training deep feedforward neural networks. In: Teh, Y.W., Titterington, M. (eds.) AISTATS 2010, PMLR, vol. 9, pp. 249--256 (2010). \url{https://proceedings.mlr.press/v9/glorot10a.html}
  
\bibitem{KingmaBa15_Adam}
Kingma, D.P., Ba, J.: Adam: A method for stochastic optimization. In: 3rd International Conference on Learning Representations (ICLR 2015), San Diego (2015). \url{https://arxiv.org/abs/1412.6980}

\bibitem{PaszkeEtAl19_PyTorch}
Paszke, A., et al.: PyTorch: An imperative style, high-performance deep learning library. In: Advances in Neural Information Processing Systems, vol. 32, Curran Associates, Inc. (2019).

\end{thebibliography}
%

\end{document}